\documentclass[aps,prl,floatfix,epsfig,twocolumn,showpacs,preprintnumbers]{revtex4}
\usepackage{amssymb}
\usepackage{graphicx}
\usepackage{amsmath}

\begin{document}

\title{Local Self--Energy Approach For Electronic Structure Calculations }
\author{N.E. Zein$^{1}$, S. Y .Savrasov$^{2}$, G. Kotliar$^{3,4}$}
\affiliation{$^{1}$RRC "Kurchatov Institute", Moscow, 123182, Russia}
\affiliation{$^{2}$Department of Physics, University of California Davis, CA 95616, USA}
\affiliation{$^{3}$Center for Material Theory, Department of Physics and Astronomy,
Rutgers University, Piscataway, New Jersey, 08854, USA}
\affiliation{$^{4}$\'{E}cole Polytechnique, 91128 Palaiseau Cedex, France}
\date{\today }

\begin{abstract}
Using a novel self---consistent implementation of Hedin's GW perturbation
theory we calculate space and energy dependent self--energy for a number of
materials. We find it to be local in real space and rapidly convergent on
second-- to third-- nearest neighbors. Corrections beyond GW are evaluated
and shown to be completely localized within a single unit cell. This can be
viewed as a fully self consistent implementation of the dynamical mean field
theory for electronic structure calculations of real solids using a
perturbative impurity solver.
\end{abstract}

\maketitle

\draft

The construction of a controlled practical approximation to the many body
problem of solid state physics is a long sought goal. Controlled
approximations are important because the accuracy of the results can be
improved in a systematic way. This goal has been achieved in quantum
chemistry by the configuration interaction (CI) method. CI can be thought as
a controlled approximation that becomes more accurate as two factors are
increased: a) the number of configurations (i.e. Slater determinants) kept
and b) the size of the basis used to represent the one-- particle orbitals
which are used to represent the configurations. Dynamical mean field theory
(DMFT) and its cluster extensions (C--DMFT) \cite{DMFTreview1}\cite%
{DMFTreview2} merge CI ideas with band structure methods. They allow us to
tackle the problem of periodic infinite systems.

The central goal is the computation of the one--electron Greens function,
(its Fourier transform can be measured via photoemission and inverse
photoemission spectroscopy), $G(\mathbf{r},\mathbf{r}^{\prime },\omega ),$
and the self--energy $\Sigma (\mathbf{r},\mathbf{r}^{\prime },\omega )$. At
the same time, following Hedin \cite{Hedin}, one introduces the effective or
screened interaction $W(\mathbf{r},\mathbf{r}^{\prime },\omega )$. The
solution of the full many--body problem can be formulated as the
extremization of a functional $L[G,W]=\mathrm{Tr}\ln G+\mathrm{Tr}\ln
W-\Sigma G-\Pi W+\Phi \lbrack G,W].$ It is defined as the Legendre transform
of thermodynamic potential with respect to non--interacting Green function
and bare Coulomb interaction \cite{Chitra}. It strongly resembles the
Luttinger--Ward functional\cite{LW} but has extremum both in the
self--energy $\Sigma $ and polarizability $\Pi $, which plays the role of
the self--energy for $W$.

\begin{figure}[tbp]
\includegraphics*[height=1.7in]{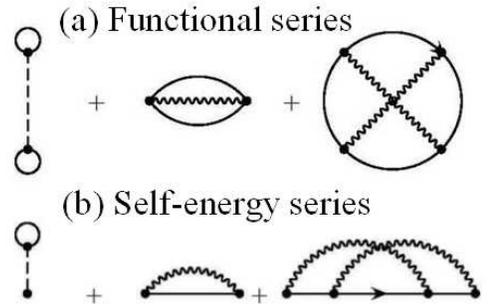}
\caption{Diagrammatic representation of the interaction energy functional $%
\Phi $ (a) leading to the so called Hedin perturbation series for the
self--energy (b). First term in the expansion is the Hartree contribution,
second represents GW diagram, third is the correction to GW.}
\label{DIAG}
\end{figure}

The interaction functional $\Phi \lbrack G,W]$ is then expanded in a
perturbative series. The first few graphs are shown in Fig \ref{DIAG}(a) and
corresponds to the Hartree, the GW, and the first correction beyond GW.
Variations of $\Phi $ over $G$ and $W$ give us $\Sigma $ and $\Pi $. For the
self--energy these diagrams are given in Fig \ref{DIAG}(b). 
To solve the corresponding Dyson equations numerically one introduces a
basis set and corresponding expansions for the self energies polarizations
and effective interactions. Cluster DMFT ideas truncate the functional $\Phi 
$, $\Sigma $ and $\Pi $ by setting its variables, i.e. the Greens functions,
equal to zero beyond a given range $R$. When $R$ is one lattice spacing we
have the highly successful single site DMFT, as the range $R$ increases the
approximation converges to the solution of the full many--body problem. In
this paper we address the central problem of determining the minimal range
that is needed to obtain accurate results for various materials.

There are three different parameters that need to be increased to achieve
convergence, a) the size of the basis set $L_{max},$ b) the order of the
perturbation theory kept $n_{max},$ c) the range of the graphs $R_{max}$
which needs to be kept to obtain accurate approximation. $R_{max}$ depends
on $L$ and $n$. We do not consider in this paper the important issue of
convergence as a function of $L_{max}$ as well as the dependence of the
range of the type of basis set chosen. Instead we make the choice of a
minimal basis set and focus on the issues of convergence as a function of $n$
and $R$.

Keeping $R_{max}$ equal to one lattice spacing and $n_{max}=\infty $ results
in the $L_{max}$ orbitally degenerate single site DMFT approximation.
Keeping $R_{max}=\infty $ and $n=1$ corresponds to the famous GW
approximation\cite{Hedin,Arya1,Hanke,Louie,Godby}. $R_{max}=1$ and $%
n_{max}=1 $ is reduced to the local GW approximation introduced by Zein and
Antropov \cite{Zein}, as an approximation to accelerate the convergence of
the GW method. Keeping $R_{max}(n=1)=\infty $ , $R_{max}(n>1)=1$ and $%
n_{max}=\infty $ constitutes the GW+DMFT approximation\cite%
{Tsvelik,Zein,Georges,SunGW}. Until now, this approximation has only been
fully implemented in the context of model Hamiltonians \cite{SunGW} and its
more realistic implementations \cite{Georges} still contain adjustable
parameters such as the double counting correction.

In this work we present a real--space cluster implementation of the GW
technique which allows us to monitor directly the locality of the
self--energy in the real space. In addition, we evaluate corrections beyond
the GW and answer the question of their convergency with respect to the
cluster size. We show that the most non--local are the contributions from
the diagrams with one loop. It was recognized early on\cite{Khodel} that the
higher the order of the diagram, the more local it is because "crossing"
integration over internal wavevectors increases the role of large momentum
leading to the locality in real space. We show here that in many real
solids, the truncation of diagrams beyond one loop to the range of one
lattice constant is already very accurate.

We also investigate the smallest value of $R_{max}$ which is needed to
obtain accurate results for each value of $L$. This allows us to obtain
fully self consistent results, independent of the starting point where local
density approximation (LDA) to density functional theory (DFT) \cite%
{DFTreview} serves in many cases \cite{Arya1,Hanke,Louie,Godby}. We
establish that even in the case of semiconductors when the Coulomb
interactions have an infinite range due to lack of screening, a reasonable
small cluster produces very accurate results. The size of the cluster needed
to obtain accurate results is reduced as the order in perturbation theory
increases.

For weakly correlated systems our approach can be regarded as a trick to
simplify and accelerate the solution of the GW equations and further
perturbation corrections to it. Alternatively, our approach should be viewed
as the first fully self--consistent implementation of an ab initio cluster
DMFT method for solids (with second order perturbation theory playing the
role of impurity solver).

We discuss the results of our calculations for a number of materials such as
simple and transition metals as well as semiconductors. Especially the
latter class represents a hard case scenario for methodologies based on
local self--energy approximations due to the long range nature of its
statically screened Coulomb interaction. We focus on the electronic
structure of Si, a benchmark in the past GW studies. Some aspects of this
problem are still debated, such as the effect of higher lying energy states,
core exchange, pseudopotential vs. all--electron approximations \cite{Ku}.
We focus on the convergence of the electronic structure as a function of the
cluster size used for determination of the self--energy for a given basis
set. We also evaluate several diagrams beyond GW to examine the issue of
convergency of the whole perturbation theory with respect to the Coulomb
interaction.

Our implementation is based on the linear muffin--tin orbital (LMTO) method
for electronic structure calculations using the atomic sphere approximation
(ASA) \cite{Andersen} which has been already used in realizations of GW \cite%
{Arya1,Kotani}. The LMTO\ basis functions separate the wavevector and radial
dependences 
\begin{equation}
\chi _{\alpha }^{\mathbf{k}}(\mathbf{r})=\Phi _{\alpha }^{H}(\mathbf{r}%
)+\sum_{L}\Phi _{L}^{J}(\mathbf{r})S_{L\alpha }^{\mathbf{k}}.  \label{CHI}
\end{equation}%
Here $S_{L\alpha }^{\mathbf{k}}$ are the LMTO$\ $structure constants while $%
\Phi _{L}^{H,J}$ are linear combinations of the solutions of the radial Schr%
\"{o}dinger equation as well as their energy derivatives taken at some fixed
set of energies $\epsilon _{\nu }$ at the center of interest which are
matched continuously and differentiably to spherical Hankel (H) and Bessel
(J) functions at the muffin--tin sphere boundaries.

In order to find the matrix elements $\Sigma _{\alpha \beta }(\mathbf{k}%
,i\omega _{n})$ of the self--energy operator $\Sigma (\mathbf{r,r}^{\prime
},i\omega _{n})$ in brackets of the LMTOs (\ref{CHI}) for a set of imaginary
Matsubara frequencies $\omega _{n}=(2n+1)\pi T$ it is useful to represent
real space vectors $\mathbf{r}=\mathbf{\rho }+\mathbf{R,}$ $\mathbf{r}%
^{\prime }=\mathbf{\rho }^{\prime }+\mathbf{R}^{\prime }\mathbf{,}$ where $%
\mathbf{\rho },\mathbf{\rho }^{\prime }$ are restricted by the unit--cell, $%
\mathbf{R,R}^{\prime }$ are the lattice translations, and redenote $\Sigma (%
\mathbf{r,r}^{\prime },i\omega _{n})=\Sigma _{R}(\mathbf{\rho ,\rho }%
^{\prime },i\omega _{n})$ (due to translational invariance we can always set 
$R^{\prime }=0$)$.$ Then, $\Sigma _{\alpha \beta }(\mathbf{k},i\omega _{n})$
has the following structure 
\begin{widetext}
 
 \begin{equation}
 \Sigma _{\alpha \beta }(\mathbf{k},i\omega _{n})=\Sigma _{\alpha \beta
 }^{(HH)}(\mathbf{k},i\omega _{n})+\sum_{L}\Sigma _{\alpha L}^{(HJ)}(\mathbf{k%
 },i\omega _{n})S_{L\beta }^{\mathbf{k}}+\sum_{L}S_{L\alpha }^{\mathbf{k\ast }%
 }\Sigma _{L\beta }^{(JH)}(\mathbf{k},i\omega _{n})+\sum_{L}S_{L\alpha }^{%
 \mathbf{k\ast }}\Sigma _{LL^{\prime }}^{(JJ)}(\mathbf{k},i\omega
 _{n})S_{L\beta }^{\mathbf{k}}  \label{SIG}
 \end{equation}%
 
 where $\Sigma _{LL^{\prime }}^{(\mu \nu )}(\mathbf{k},i\omega
 _{n})=\sum_{R}e^{i\mathbf{kR}}\Sigma _{LL^{\prime }}^{(\mu \nu )}(\mathbf{R}%
 ,i\omega _{n})$ and the cluster self--energy is given by the matrix element%
 \begin{equation}
 \Sigma _{LL^{\prime }}^{(\mu \nu )}(\mathbf{R},i\omega _{n})=\int \Phi
 _{L}^{(\mu )\ast }(\mathbf{\rho })\Sigma _{R}(\mathbf{\rho ,\rho }^{\prime
 },i\omega _{n})\Phi _{L\prime }^{(\nu )}(\mathbf{\rho }^{\prime })d\mathbf{%
 \rho }d\mathbf{\rho }^{\prime }.  \label{MAT}
 \end{equation}
 \end{widetext}As we see, even if $\Sigma (\mathbf{r,r}^{\prime },i\omega
_{n})$ can be local (i.e. non--zero only when both $\mathbf{r}$ and $\mathbf{%
\ r}^{\prime }$ are in the same cell), the matrix elements $\Sigma _{\alpha
\beta }(\mathbf{k},i\omega _{n})$, Eq. ({\ref{SIG}), acquire some }$\mathbf{k%
}${--dependence through the structure constants. It is due to tails of the
basis functions extended over all space, and it is quite analogous to the }${%
\ \mathbf{k}}${--dependence of local potential matrix elements in LDA. Such }%
$\mathbf{k}${--dependence can be called "kinematical", and in the following
we will distinguish it from the dynamical }$\mathbf{k}${\ dependence
connected to the existence of }$\mathbf{R}\neq 0$ elements of $\Sigma _{R}(%
\mathbf{\rho },\mathbf{\rho }^{\prime },i\omega _{n})${\ which is the main
focus of the present work.}

\begin{table}[tbp]
\caption{Correlational contribution to the self--energy matrix element, $%
\Sigma _{c,ll}^{HH}(\mathbf{R},i\protect\omega _{0}),$ in eV as a function
of $\mathbf{R}$ for Fe, Ni, Na, Al and Si. $\mathbf{R}_{1}=a/\protect\sqrt{2}
$, $\mathbf{R}_{2}=a$, $\mathbf{R}_{3}=a\protect\sqrt{2}$ in fcc structure), 
$\mathbf{R}_{1}=a\protect\sqrt{3}/2$ $\mathbf{R}_{2}=a$, $\mathbf{R}_{3}=a%
\protect\sqrt{3}/2$ in bcc structure, $\mathbf{R}_{1}=0.5a$, $\mathbf{R}%
_{2}=0.83a$ $\mathbf{R}_{3}=a$ in diamond structure.}
\label{SIGR}%
\begin{tabular*}{\columnwidth}{@{\extracolsep{\fill}}ccccc}
\hline\hline
& $\mathbf{R}=0$ & $\mathbf{R}_{1}$ & $\mathbf{R}_{2}$ & $\mathbf{R}_{3}$ \\ 
\hline
Fe: $\Sigma _{ss}(\mathbf{R})$ & 0.70 & 0.35 & 0.16 & 0.00 \\ 
Fe: $\Sigma _{dd}(\mathbf{R})$ & 6.53 & 0.05 & 0.08 & 0.00 \\ 
Ni: $\Sigma _{ss}(\mathbf{R})$ & -0.54 & 0.05 & 0.03 & 0.00 \\ 
Ni: $\Sigma _{dd}(\mathbf{R})$ & 7.34 & 0.38 & -0.03 & 0.00 \\ 
Na: $\Sigma _{ss}(\mathbf{R})$ & -1.36 & 0.52 & 0.30 & 0.05 \\ 
Al: $\Sigma _{pp}(\mathbf{R})$ & 0.46 & 0.16 & -0.08 & 0.00 \\ 
Si: $\Sigma _{ss}(\mathbf{R})$ & 0.95 & 0.30 & 0.14 & 0.03 \\ 
Si: $\Sigma _{pp}(\mathbf{R})$ & -1.06 & 0.05 & -0.27 & -0.03 \\ \hline\hline
\end{tabular*}%
\end{table}

From Eq. (\ref{CHI}) follows that the one--electron Green function can be
represented in a factorized form 
\begin{equation}
G_{R}(\mathbf{\rho ,\rho }^{\prime },i\omega _{n})=\sum_{\mu \nu =HJ}\Phi
_{L}^{(\mu )\ast }(\mathbf{\rho })G_{LL^{\prime }}^{(\mu \nu )}(\mathbf{R}%
,i\omega _{n})\Phi _{L\prime }^{(\nu )}(\mathbf{\rho }^{\prime })
\label{GRN}
\end{equation}%
\bigskip and, as a result, the polarization operator $\Pi (\mathbf{r},%
\mathbf{r}^{\prime },i\omega _{n})$, has a similar structure. We solve the
equation for the dynamically screened interaction $W=V-V\Pi W$ on the
product of basis functions following Ref. \onlinecite{AryaGunn}. After
finding $W(\mathbf{r},\mathbf{r}^{\prime },i\omega _{n})$ the self--energy
is calculated either as $\Sigma (\mathbf{R})=G(\mathbf{R})W(\mathbf{R})$ ($%
\mathbf{R}$--space version) or as $\Sigma (\mathbf{k})=\sum_{\mathbf{k}}G(%
\mathbf{k})W(\mathbf{k+q})$ ($\mathbf{k}$--space version). All calculations
are performed on the imaginary axis. Due to large frequency behavior of the
Green function proportional to $1/i\omega _{n}$ a special care should be
taken of the direct exchange contribution to the self--energy ($\Sigma
_{x}=-GV$ ), as the sum over large $\omega _{n}$ needs to be done
analytically. The remaining portion, $\Sigma _{c}=G(V\Pi W),$ is due to
correlations, and the sum over internal frequencies is rapidly convergent.
Finally, in order to obtain the electronic spectrum for real frequencies we
analytically continue the Green function using Pade's approximation.

\begin{table}[tbp]
\caption{ Polarizability $\Pi _{ij}(\mathbf{R,}\protect\omega =0)$ as a
function of cluster size $\mathbf{R}$ as well as $\Pi _{ij}(\mathbf{k}=0%
\mathbf{,}\protect\omega =0)$ for Si ($ij$=1,2 numerates Si atoms, $ij$=3,4
numerates empty spheres) . $\sum \Pi $ shows how the sum rule $\Pi
(k\rightarrow 0,\protect\omega \equiv 0)\rightarrow 0$ is fulfilled in both $%
\mathbf{R}$--space and $\mathbf{k}$--space calculations.}
\label{TPIR}%
\begin{tabular*}{\columnwidth}{@{\extracolsep{\fill}}cccccc}
\hline\hline
& $\Pi _{11}$ & $\Pi _{12}$ & $\Pi _{13}$ & $\Pi _{14}$ & $\sum \Pi $ \\ 
\hline
$\mathbf{R}=0$ & 5.34 & 0 & 0 & 0 & 14.78 \\ 
$\mathbf{R}_{1}$ & 4.98 & -2.60 & -0.55 & 0.21 & 6.76 \\ 
$\mathbf{R}_{2}$ & 4.98 & -2.67 & -1.00 & -0.09 & 3.45 \\ 
$\mathbf{R}_{3}$ & 4.98 & -3.33 & -1.159 & -0.21 & 0.81 \\ 
$\mathbf{R}_{4}$ & 4.96 & -3.36 & -.14 & -0.23 & 0.41 \\ 
$\mathbf{k}$--space & 4.88 & -3.40 & -1.21 & -0.26 & 10$^{-11}$ \\ 
\hline\hline
\end{tabular*}%
\end{table}

We perform self--consistent GW calculations for Fe, Ni, Na, and Si using our
newly developed cluster algorithm and obtain the self--energies in real
space. To discuss the results of these calculations, Table \ref{SIGR} lists
the diagonal matrix elements of the correlational part of the self--energy $%
\Sigma _{c,ll}^{(HH)}(\mathbf{R},i\omega _{0}),$ (\ref{CHI}), as a function
of $\mathbf{R}$ for the value of $\omega _{0}=\pi T=\pi 400K$. (In case of
magnetic ground state $\Sigma _{c,ll}^{HH}(i\omega _{0})$ is for majority
spins.) From Table \ref{SIGR} it follows that $\Sigma _{c,dd}(\mathbf{R}=0)$
dominates in transition metals and falls off very quickly in nearest
neighbors, because of both small overlap between d--orbitals and large
screening at small energies. The $\Sigma _{c,ss}(\mathbf{R})$ in these
metals falls off more gradually but its value is negligible in comparison
with $d$. It explains why recently developed one--site approach \cite{Zein}
is successful in this case. In simple metals like Na and Al only $\Sigma
_{s} $ and $\Sigma _{p}$ are significant, they fall off gradually, but
nevertheless on the third coordination sphere they become small as compared
to $\Sigma (\mathbf{R}=0)$, even if they are multiplied by a number of
nearest neighbors at this sphere. In Si, both $\Sigma _{c,ss}(\mathbf{R})$
and $\Sigma _{c,pp}(\mathbf{R})$ fall off pretty slowly but nevertheless
become very small at $\mathbf{R}$'s comparable with the size of the unit
cell in accord with the conclusions reached in the pioneering work \cite%
{Louie}.

A proper approximation of the polarizability $\Pi $ of a semiconductor,
requires going beyond the local approximation in order to fulfill the f sum
rule $\Pi (\mathbf{k}\rightarrow 0,\omega \equiv 0)\rightarrow 0.$ This
requires cancellations between local and non--local terms. In Table \ref%
{TPIR} we show how this sum rule is fulfilled for Si, where values of $\Pi
_{ij}(\mathbf{R},\omega \equiv 0)$ ($ij$ numerate atoms in the unit cell)
are listed for several $\mathbf{R}$'s. As we utilize the atomic sphere
approximation, in order to reach close packing we consider two Si atoms ($ij$%
=1,2) and two empty spheres ($ij$=3,4). Table \ref{TPIR} also lists the
matrix elements $\Pi _{ij}(\mathbf{k}=0,\omega \equiv 0)$ obtained by the $%
\mathbf{k}$--space calculation as well as sums, $\sum \Pi ,$ over the atomic
coordinates, which show how the sum rule is fulfilled. We see that while the 
$\mathbf{k}$ --space version automatically leads to obeying the sum rule,
the $\mathbf{R}$ --space version needs cluster sizes extended up to four
nearest neighbors.

\begin{figure}[tbp]
\includegraphics*[height=2.4in]{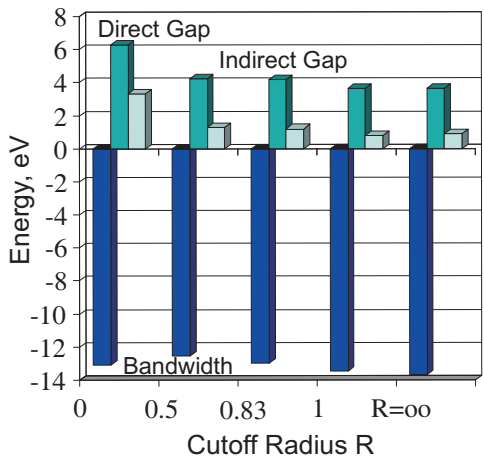}
\caption{Direct and indirect energy gaps as well as valence bandwidth for Si
in eV calculated by varying the size of the cluster used when evaluating
correlational part of the self--energy within the GW method.}
\label{ELSI}
\end{figure}

We now investigate the convergency of the electronic structure of Si using
the cluster GW method. Both the direct and indirect energy gaps as well as
the valence band width are calculated by varying the size of the cluster
used when evaluating correlational part of the self--energy. The behavior of
the electronic structure as a function of the cluster size is schematically
shown on Fig \ref{ELSI}. They indicate that the non--locality of the
self--energy in Si approximately extends up to a third coordination sphere.

We have evaluated the first self--energy correction beyond GW, $\Delta
\Sigma (\mathbf{R})$, for Si as a function of the cluster size. Our obtained
on--site values for s and p electrons are as follows: $\Delta \Sigma _{s}(0)$
=0.22 eV and $\Delta \Sigma _{p}(0)$ =--0.14 eV. The self--energies at first
and second coordination sphere are given by: $\Delta \Sigma _{s}(\mathbf{R}%
_{1})$ =0.002 eV, $\Delta \Sigma _{p}(\mathbf{R}_{1})$ =--0.002 eV, $\Delta
\Sigma _{s}(\mathbf{R}_{2})$ =0.0003 eV, $\Delta \Sigma _{p}(\mathbf{R}_{2})$
=--0.0005 eV. We see that the correction to GW is completely local and
permits us to calculate it in the real space which much less
time--consuming. It is also interesting to note that the energy dependence
of this correction exists but at the same scale as in $\Sigma _{GW}.$

We finally would like to address a highly interesting question on the
accuracy of the GW calculations in predicting the energy gaps in various
semiconductors and insulators. We have performed such calculations with and
without imposing the self--consistency for the Green function which in the
latter case corresponds to the LDA Green function used in evaluating the
interactions and the self--energies. We refer to these calculations as
self--consistent (SC) and "first shot" (FS) ones. We include the core
exchange effects, whose importance has been recently pointed out \cite{Ku}.
The results of these studies are presented in Table \ref{GAPS} where we list
the obtained energy gaps for a whole series of materials such as C, Si, MgO,
and AlAs. The available previous calculations and experimental data are also
listed in this table for comparison. Despite the possible inaccuracies
introduced by the use of a limited basis set, which does not include very
high--energy states, we see that our non--self--consistent calculations are
in good agreement with the published data. The self--consistency is not only
of conceptual importance, since it makes the solution of the problem
independent on the starting point, but can also affect the quality of the
results in some systems such as AlAs.

In conclusion we have developed a self-consistent cluster DMFT methodology
which allows us to monitor the locality of the self--energy in the real
space. As a first application of the method we evaluated first contribution
in Hedin's perturbation series for the self--energy in Si and found it to be
completely local. Our calculated energy gap values for a number of
semiconductors and insulators are found in good agreement with experiment.
The fast convergence in real space demonstrated in this paper simplifies and
accelerates the GW approximation which is useful in many solids. Our
approach is also the ideal starting point for resuming the perturbation
theory at the local level using non--perturbative solvers as was done in
Ref. \cite{SunGW}. This step will allow us to study very strongly correlated
systems completing the first principles C--DMFT program for solids.

\begin{table}[tbp]
\caption{Comparison between calculated energy gaps (eV)\ in semiconductors
and insualtors using non--self--consistent "first shot" (FS) and
self--consistent (SC) GW methods obtained in the present work, results of
other available GW calculations and experiment.}
\label{GAPS}%
\begin{tabular*}{\columnwidth}{@{\extracolsep{\fill}}cccccc}
\hline\hline
& FS (this work) & SC (this work) & FS & SC & Exp. \\ \hline
C & 5.00 & 5.02 & 4.92$^{a}$ & -- & 5.48$^{b}$ \\ 
Si & 0.86 & 1.10 & 0.85$^{a}$ & 1.05$^{c}$ & 1.17$^{b}$ \\ 
MgO & 8.00 & 5.90 & 8.3$^{d}$ & -- & 7.8$^{d}$ \\ 
AlAs & 1.33 & 1.90 & 1.05$^{a}$ & -- & 2.24$^{b}$ \\ \hline\hline
\end{tabular*}
$^{a}$Ref.\onlinecite{Godby};$^{b}$Ref.\onlinecite{Landolt};$^{c}$Ref.%
\onlinecite{Ku}; $^{d}$Ref. \onlinecite{Fu};$^{e}$Ref.\onlinecite{MgOexp};
\end{table}

This work was supported by the Council for the Support of Leading Scientific
Schools of Russia under grant NS--1572.2003.2. Support from NSF--ITR Grants
No. 0312478, 0342290 and DOE Computational Material Science network is
gratefully acknowledged together with NSF DMR Grants No. 0096462, 02382188,
and DOE Grants DEFG02--99ER45761, LDRD--DR 200030084. GK is supported by the
Blaize Pascal Chair of the Foundation de l'Ecole Normale.


\begin{thebibliography}{99}
\bibitem{DMFTreview1} For a review, see, A. Georges, G. Kotliar, W. Krauth,
and M.~Rozenberg, Rev. Mod. Phys. \textbf{68}, 13 (1996).

\bibitem{DMFTreview2} For a review, see, G. Kotliar, S. Savrasov, K. Haule,
V. Oudovenko, C. Marianetti, and O. Parcolett, cond--mat/0511085.

\bibitem{Hedin} L. Hedin, Phys. Rev. \textbf{139}, A796 (1965).

\bibitem{Chitra} C--. O. Almbladh, U. von Barth, R. van Leeuwen, Int. J.
Mod. Phys. B \textbf{13}, 535 (1999); R. Chitra, G. Kotliar Phys. Rev. B%
\textbf{\ 63}, 115110 (2001).

\bibitem{LW} J. M. Luttinger, J. C. Ward Phys. Rev. \textbf{118}, 1417
(1960).

\bibitem{Arya1} For a review, see, F. Aryasetiawan, O. Gunnarson Rep. Prog.
Phys. \textbf{61}, 237 (1998).

\bibitem{Hanke} G. Strinati, H. J. Mattausch, and W. Hanke, Phys. Rev. Lett. 
\textbf{45}, 290-294 (1980).

\bibitem{Louie} M. S. Hybertsen and S. G. Louie, Phys. Rev. Lett. \textbf{55}
, 1418 (1985).

\bibitem{Godby} R.W. Godby, M. Schl\"{u}ter, and L. J. Sham, Phys. Rev.
Lett. \textbf{56}, 2415 (1986).

\bibitem{Zein} N. E. Zein and V. P. Antropov, Phys. Rev. Lett. 89, 126402
(2002).

\bibitem{Tsvelik} G. Kotliar and S. Y. Savrasov, in \textit{New Theoretical
approaches to strongly correlated systems}, edited by A. M. Tsvelik, (Kluwer
Academic Publishers, the Netherlands, 2001), p. 259, (available in
cond--mat/020824).

\bibitem{Georges} S. Biermann, F. Aryasetiawan, and A. Georges, Phys. Rev.
Lett. \textbf{90}, 086402 (2003).

\bibitem{SunGW} P. Sun, G. Kotliar, Phys. Rev. Lett. \textbf{92}, 196402
(2004).

\bibitem{Khodel} V.A. Khodel, E. E. Saperstein, Phys. Lett. B \textbf{36},
429 (1971).

\bibitem{DFTreview} For a review, see, \textit{e.g.}, \textit{Theory of the
Inhomogeneous Electron Gas}, edited by S. Lundqvist and S. H. March (Plenum,
New York, 1983).

\bibitem{Ku} W. Ku and A. G. Eguiluz, Phys. Rev. Lett. \textbf{89}, 126401
(2002).

\bibitem{Andersen} O. K. Andersen, Phys. Rev. B \textbf{12}, 3050 (1975).

\bibitem{Kotani} S. V. Faleev, M. van Schilfgaarde, and T. Kotani, Phys.
Rev. Lett. \textbf{93}, 126406 (2004).

\bibitem{AryaGunn} F. Aryasetiawan, O. Gunnarson Phys. Rev. B \textbf{49,}
16214 (1994).

%\bibitem{Jackiw} J.M. Cornwall, R.Jackiw, E. Tomboulis Phys. Rev. \textbf{D10%
%}, 2428 (1974).

\bibitem{Landolt} Physics of Group IV Elements and III-V Compounds,
Landolt-Bornstein,. Numerical Data and Functional Relationships in Science
and Technology, ed. by O. Madelung, M. Schultz, H.Weiss (Springer,N.-Y.
1982) Group III, v17a.

\bibitem{Fu} A.Yamasaki, T.Fujiwara Phys.Rev B \textbf{66}, 245108 (2002).

\bibitem{MgOexp} R.C. Whited, C.J.Flaten, W.C. Walker Solid St. Comm. 
\textbf{13},1903 (1973).

%\bibitem{Sun} P. Sun, G. Kotliar Phys. Rev. \textbf{B 66}, 085120 (2002)

%\bibitem{Taylor} A. Holas, P.K. Aravind, K.S. Singwo Phys. Rev. \textbf{B20}, 4912 (1979).

%\bibitem{godby} A.Schindlmayr, T.J. Pollehn, R.W. Godby
%Phys. Rev. {\bf B58}, 12684 (1998)

%\bibitem{Arya3} F. Aryasetiawan Phys. Rev. \textbf{B45}, 13051 (1992)

%\bibitem{khodel2} V.A. Khodel, E.E. Saperstein Phys. Rep. 92, 183 (1982)
\end{thebibliography}
\end{document}